# A Theory for Analysis of Pulse Electromagnetic Radiation

Gaobiao Xiao, Shanghai Jiaotong University

gaobiaoxiao@sjtu.edu.cn

**Abstract**—A theory for analyzing the radiative and reactive energies for pulse radiators in free space is presented. With the proposed definition of reactive energies and radiative energies, power balance at arbitrarily chosen observation surfaces are established, which intuitively shows that the Poynting vector contains not only the power flux density associated with the radiative energies, but also the influence of the fluctuation of the reactive energies dragging by the sources. A new vector is defined for the radiative power flux density. The radiative energies passing through observation surfaces enclosing the radiator are accurately calculated. Numerical results verifies that the proposed radiative flux density is more proper for expressing the radiative power flux density than the Poynting vector.

## I. Introduction

The electromagnetic radiation problems have been intensively investigated for more than a hundred years. It is a little bit strange that there is still no widely accepted formulation for evaluating the stored reactive energies and Q factors of radiators[1]-[14]. The main difficulty may come from the fact that there is no clear definition in macroscopic electromagnetic theory for the reactive electromagnetic energy. It is commonly known in classical charged particle theory that the fields associated with charged particles can be divided into self fields and radiative fields[15][16]. The self fields include the Coulomb fields and the velocity fields, carrying self energies, also referred to as Schott energy in some literatures [17]-[18][19]. The radiative fields are generated by acceleration of charged particles, emitting radiative energies to the surrounding space. The self fields/energies are considered to be attached to the charged particles, or simply speaking, they show up with the charged particles and disappear with the charged particles. On the contrary, after being radiated by the charged particles, the radiative fields/energies will depart from the sources and propagate to the remote infinity. They exist even after their generating sources disappeared and can couple with other sources they encountered in their journey. Although it is natural to consider that the reactive energies in macroscopic electromagnetics is similar to the self energies or the Schott energy, no successful attempt has been found or well accepted to handle the reactive energies in this manner. No expression for reactive energies is established in macroscopic electromagnetics that can be derived rigorously from the self fields of charged particles.

Poynting vector is widely considered as the electromagnetic power flux density[20]. Poynting Theorem describes the relationship between the Poynting vector, the varying rate of the electromagnetic energy densities, and the work rate done by the exciting source. It provides an intuitive description of the propagation of electromagnetic energy. However, interpreting the Poynting vector as the electromagnetic power flux density has always been controversial [21]-[40], and some researchers have pointed out that Poynting Theorem may have not been used in the correct way in some situations[41][42]. However, most of these opinions have been ignored because of the great success of the wide application of Poynting Theorem and Poynting vector.

It is known that the Poynting Theorem is not convenient to use for evaluating the reactive energies stored by radiators in an open space [5][13], which has been investigated for decades. For harmonic fields, the total electromagnetic energy obtained by integrating the conventional energy densities of $(0.5\mathbf{D}\cdot\mathbf{E})$ and $(0.5\mathbf{B}\cdot\mathbf{H})$ over the whole space is infinite because the conventionally defined electric and magnetic energy densities generally account for the total fields consisting of the radiative fields and the reactive fields. The radiative energy occupies the whole space and is infinitely large [14]. Some researchers suggested that those fields associated with the propagating waves should not contribute to the stored reactive energies, and the results can become finite by subtracting from the energy density an additional term associated with the radiation power. However, it is not easy to give a general definition for the term because the propagation patterns are quite different for different radiators [1][5].

Practically, if we check the classical charged particle theory, the Larmor's formula for the radiative power of an accelerated charged particle can be derived from the corresponding Poynting vector with contribution from the radiative fields only[15]-[17][42].

Based on these observations, the macroscopic electro- magnetic radiation issue is revisited and a new energy/power balance equation at a certain instant time is proposed, which gives an intuitive and reasonable suggestion that Poynting vector is not the radiative power flux density.

It is not the aim of this paper to provide a rigorous proof to support that the reactive energies in the macroscopic electromagnetics are exactly the self energies or the Schott energy in the classical charged particles. Instead, a definition for reactive electromagnetic energies is proposed based on the hypothesis that the reactive energies in the macroscopic electromagnetics bear the same characteristics as the self energies: (1) they are attached to the sources, appear/disappear with the sources simultaneously; (2) the definition is in consistent with the static energies associated with Coulomb fields; (3) the reactive energies do not propagate like radiative energies, but their fluctuation may propagate at the light velocity in free space just like the radiative fields. A theory is proposed based on these considerations, in which the radiative energies and the reactive energies can be separated. As a consequence, the Poynting vector is divided into two vectors. One vector accounts for the radiative power flux density and the other vector accounts for the fluctuation of the reactive energies. It has to mention that the theory is nonrelativistic and based on macroscopic Maxwell theory. The proposed theory enjoys success in interpreting the radiation process of a Hertzian dipole, providing results exactly in agreement with those obtained using the well-established Chu's circuit model for the dipole. The theory is also supported by numerical examples.

## II. Formulations for Reactive and Radiative Energies

In the proposed theory, the reactive electric and magnetic energy of a radiator are defined with

$$W_\rho(t) = \int_{V_s} \frac{1}{2}\rho(\mathbf{r}',t)\phi(\mathbf{r}',t)d\mathbf{r}' = \int_{V_s} w_\rho(\mathbf{r}',t)d\mathbf{r}' \tag{1}$$

$$W_J(t) = \int_{V_s} \frac{1}{2}\mathbf{A}(\mathbf{r}',t)\cdot\mathbf{J}(\mathbf{r}',t)d\mathbf{r}' = \int_{V_s} w_J(\mathbf{r}',t)d\mathbf{r}' \tag{2}$$

where the scalar potential $\phi$ and vector potential $\mathbf{A}$ evaluated at the observation point $\mathbf{r}$ and the instant of time $t$ are defined in their usual way,

$$\phi(\mathbf{r},t) = \int_{V_s} \frac{\rho(\mathbf{r}',t')}{4\pi\varepsilon_0 R} d\mathbf{r}' \tag{3}$$

$$\mathbf{A}(\mathbf{r},t) = \mu_0 \int_{V_s} \frac{\mathbf{J}(\mathbf{r}',t')}{4\pi R} d\mathbf{r}' \tag{4}$$

In the above equations, $\rho(\mathbf{r}',t')$ and $\mathbf{J}(\mathbf{r}',t')$ are the charge density and current density at source point $\mathbf{r}' \in V_s$ and retarded time $t' = t - R/c$, in which c is the light velocity and $R = |\mathbf{r} - \mathbf{r}'|$ is the distance. $\mu_0$ and $\varepsilon_0$ are respectively the permittivity and permeability in free space. The potentials have to satisfy the Lorentz Gauge, and their reference zero points are at the infinity.

The reactive electromagnetic energy is the sum of the reactive electric energy and the reactive magnetic energy,

$$W_{react}(t) = \int_{V_s} \left(\frac{1}{2}\rho\phi + \frac{1}{2}\mathbf{A}\cdot\mathbf{J}\right)d\mathbf{r}' \tag{5}$$

It can be readily checked that the reactive energies defined in (3), (4) and (5) are attached to their sources, i.e., they show up together with their sources and disappear with their sources. For static electromagnetic fields, they are exactly the stored electro- magnetic energies associated with the source. Notice that no other definition for energies in the macroscopic electro- magnetic theory bears this property.

For time varying fields, the reactive electric energy and the reactive magnetic energy may become negative because of the retardation. For example, the direction of the vector potential may be negative to that of the current, which is sometimes encountered in loop current sources. However, the total reactive energies for bounded sources have to be positive if we choose the infinity as the zero reference points for the potentials. Therefore, in this theory the reactive electromagnetic energy is defined combining the two reactive energies together and treated as a whole.

The Poynting Theorem correctly describes the relationship between the work rate done by the source, the total electromagnetic energy in region $V_a \supseteq V_s$ containing the source, and the total electromagnetic power flux crossing the boundary $S_a$ of the region,

$$-\int_{V_s} \mathbf{E}\cdot\mathbf{J}d\mathbf{r}' = \frac{\partial}{\partial t}\int_{V_a}\left(\frac{1}{2}\mathbf{D}\cdot\mathbf{E} + \frac{1}{2}\mathbf{B}\cdot\mathbf{H}\right)d\mathbf{r}' + \oint_{S_a} \mathbf{S}\cdot\hat{\mathbf{n}}dS \tag{6}$$

where the Poynting vector $\mathbf{S} = \mathbf{E}\times\mathbf{H}$ is conventionally regarded as the power flux density, like in the antenna society.

From Maxwell equation, we can derive equations,

$$\begin{cases} \frac{1}{2}\mathbf{D}\cdot\mathbf{E} = \frac{1}{2}\phi\rho - \frac{1}{2}\nabla\cdot(\phi\mathbf{D}) - \frac{1}{2}\mathbf{D}\cdot\frac{\partial\mathbf{A}}{\partial t} \\ \frac{1}{2}\mathbf{B}\cdot\mathbf{H} = \frac{1}{2}\mathbf{A}\cdot\mathbf{J} + \frac{1}{2}\mathbf{A}\cdot\frac{\partial\mathbf{D}}{\partial t} + \frac{1}{2}\nabla\cdot(\mathbf{A}\times\mathbf{H}) \end{cases} \tag{7}$$

Substituting (7) into (6) and reorganizing it gives

$$-\int_{V_s} \mathbf{E} \cdot \mathbf{J} d\mathbf{r}' = \frac{\partial}{\partial t} \int_{V_a} \left( \frac{1}{2}\phi\rho + \frac{1}{2}\mathbf{A}\cdot\mathbf{J} + \frac{1}{2}\mathbf{A}\cdot\frac{\partial \mathbf{D}}{\partial t} - \frac{1}{2}\mathbf{D}\cdot\frac{\partial \mathbf{A}}{\partial t} \right) d\mathbf{r}' + \oint_{S_a} \left[ \mathbf{E}\times\mathbf{H} + \frac{1}{2}\frac{\partial}{\partial t}(\mathbf{A}\times\mathbf{H} - \phi\mathbf{D}) \right] \cdot \hat{\mathbf{n}} dS \quad (8)$$

The integrand of the first term in the RHS can be interpreted as the total energy stored in $V_a$, which consists of the reactive energy and the radiative energy. Since the first two terms are defined as the reactive energies, it is natural to interpret the other two terms as the radiative energies temporally existing in the volume. We define explicitly the radiative energy in $V_a$ as

$$W_{rad} = \int_{V_a} \left( \frac{1}{2}\mathbf{A}\cdot\frac{\partial \mathbf{D}}{\partial t} - \frac{1}{2}\mathbf{D}\cdot\frac{\partial \mathbf{A}}{\partial t} \right) d\mathbf{r}' = \int_{V_a} w_{rad}(\mathbf{r}',t) d\mathbf{r}' \quad (9)$$

in which the radiative energy density is defined by,

$$w_{rad}(\mathbf{r},t) = \frac{1}{2}\mathbf{A}\cdot\frac{\partial \mathbf{D}}{\partial t} - \frac{1}{2}\mathbf{D}\cdot\frac{\partial \mathbf{A}}{\partial t} \quad (10)$$

Integrating the LHS of (6) gives the total work done by the source

$$W_{exc}(t) = -\int_{-\infty}^{t} \int_{V_s} \mathbf{E}(\mathbf{r}',\tau)\cdot\mathbf{J}(\mathbf{r}',\tau) d\mathbf{r}' d\tau \quad (11)$$

The integrand of the second term in the RHS of (8) is a flux density. We introduce a new vector for it,

$$\mathbf{S}_{rad} = \mathbf{E}\times\mathbf{H} - \frac{\partial}{\partial t}\left( \frac{1}{2}\mathbf{H}\times\mathbf{A} + \frac{1}{2}\phi\mathbf{D} \right) \quad (12)$$

(8) can then be rewritten in a compact form,

$$\frac{\partial W_{exc}}{\partial t} - \frac{\partial W_{react}}{\partial t} = \frac{\partial W_{rad}}{\partial t} + \oint_{S_a} \mathbf{S}_{rad}\cdot\hat{\mathbf{n}} dS \quad (13)$$

In the LHS of (13), the increasing rate of the reactive energy is subtracted from the total work rate by the source. By equating to it, the right-hand side of (13) can naturally be interpreted as the radiative energies. The first term of the RHS represents the increasing rate of the radiative energy in the volume, the second term represents the radiative flux. Hence, it is reasonable to interpret the vector $\mathbf{S}_{rad}$ as the radiative power flux density. Integrating it on the observation surface $S_a$ yields the total power crossing the surface at an instant of time $t$,

$$P_{Srad}(t) = \oint_{S_a} \mathbf{S}_{rad}\cdot\hat{\mathbf{n}} dS \quad (14)$$

Define a new vector,

$$\mathbf{S}_{react} = \frac{\partial}{\partial t}\left( \frac{1}{2}\mathbf{H}\times\mathbf{A} + \frac{1}{2}\phi\mathbf{D} \right) \quad (15)$$

The Poynting vector can then be divided into two parts,

$$\mathbf{S} = \mathbf{S}_{rad} + \mathbf{S}_{react} \quad (16)$$

$\mathbf{S}_{react}$ is dependent on the fields and the potentials. It is not a real power transportation by the propagating waves, but rather reflects the influence of the fluctuation of the reactive energies. For time varying sources, their reactive energies may be dragged back and forth by the motion of sources, causing a pseudo power flux crossing the observation surface $S_a$, as was observed in [42]. Judging from the expressions for the retarded potentials, it can be conformed that the fluctuation of the reactive energies also propagates at the light velocity in free space.

It is straightforward to prove that[44]

$$\lim_{r\to\infty}\left\{ \oint_{S_a} \mathbf{S}_{rad}\cdot\hat{\mathbf{n}} dS \right\}_{av} = \lim_{r\to\infty}\left\{ \oint_{S_a} \mathbf{S}\cdot\hat{\mathbf{n}} dS \right\}_{av} \quad (17)$$

At remote infinity, the Poynting vector approximately equals the radiative power flux density.

## III. Radiation of Pulse Sources

Assume that there is a symmetrical source in a sphere with radius $r_s$ in time period of $0 \leq t \leq T$. All radiative fields are spherical waves due to the symmetry. For $0 \leq t \leq T$, on the one hand, the source will induce self fields and emit radiative fields. On the other hand, they will interact with the surrounding fields generated by them at retarded time $t'$, similar to charged particles. The interaction with fields may possibly turn the radiator into an absorber at some instant of time. The radiative fields exist in the sphere $V_{rad}$, which expands with the propagation of the radiative waves. At $t = T$, the radius of $V_{rad}$ is $(r_s + cT)$.

For $t > T$, the reactive energy disappears simultaneously with the source. The region contains the radiative fields becomes a spherical shell with thickness of $r_s$. Denote the boundary of the region occupied by the radiative fields as $S_{rad}$. It has an outer and an inner boundary when $t > T$, as shown in Fig.1.

If we put the observation surface $S_a$ in the region outside of $S_{rad}$, then the surface integral in (13) is zero. Integrating both sides of (13) from $-\infty$ to t yields

$$W_{rad}(t) = W_{exc}(t) - W_{react}(t) \tag{18}$$

Since the pulse source exist only during $0 \leq t \leq T$, the reactive energy is zero for $t < 0$ and $t > T$. The total radiative energy is a constant, equal to the total work done by the source,

$$W_{rad}(t) = -\int_0^T \int_{V_a} \mathbf{E} \cdot \mathbf{J} d\mathbf{r}' d\tau = W_{exc}(T), \quad \text{for } t > T \tag{19}$$

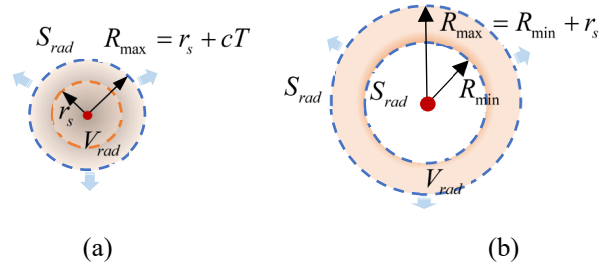

(a)            (b)

Fig.1 Region occupied by radiative fields. (a) $0 \leq t \leq T$, (b) $t > T$.

Now consider a fixed observation surface $S_a$ containing the source region $V_s$. The total radiative energy crossing the observation surface can be calculated by integrating the radiative power crossing it,

$$W_{rad} = \int_{t_{\min}}^{t_{\max}} \oint_{S_a} \mathbf{S}_{rad} \cdot \hat{\mathbf{n}} dS dt = \int_{V_a} w_{rad} d\mathbf{r}' = W_{exc}(T) \tag{20}$$

where $t_{\min} = R_{\min}/c$, $t_{\max} = R_{\max}/c$, $R_{\min}$ and $R_{\max}$ are respectively the smallest and largest distance between the source and the observation surface. As indicated in (20), the radiative energy can be directly calculated in two ways. One is to determine the region $V_{rad}$ occupied by the radiative fields at time t, and perform a volume integration with (10). The other is to choose a fixed observation surface, and accumulate the radiative power crossing it. The first method may be not efficient as it is quite difficult to accurately determine the radiative region of ordinary radiators. For the second method, some analytical expressions can be used to accurately and efficiently calculate the radiative fields on observation surfaces with regular shapes, such as spherical or cuboidal surfaces[43].

For point sources like Hertzian dipoles, their stored reactive energies are finite only after excluding a small sphere $V_a$ containing them. In this case, the reactive energies cannot be directly calculated using (5). However, they can be calculated using the fields and potentials according to the following relationships derived from Maxwell equations [44],

$$\int_{V_s} \frac{1}{2} \phi \rho d\mathbf{r}' = \int_{V_\infty - V_a} \left( \frac{1}{2} \mathbf{D} \cdot \mathbf{E} + \frac{1}{2} \mathbf{D} \cdot \frac{\partial \mathbf{A}}{\partial t} \right) d\mathbf{r}' + \oint_{S_\infty} \left( \frac{1}{2} \phi \mathbf{D} \right) \cdot \hat{\mathbf{n}} dS \tag{21}$$

$$\int_{V_s} \frac{1}{2} \mathbf{A} \cdot \mathbf{J} d\mathbf{r}' = \int_{V_\infty - V_a} \left( \frac{1}{2} \mathbf{B} \cdot \mathbf{H} - \frac{1}{2} \mathbf{A} \cdot \frac{\partial \mathbf{D}}{\partial t} \right) d\mathbf{r}' + \oint_{S_\infty} \left( \frac{1}{2} \mathbf{H} \times \mathbf{A} \right) \cdot \hat{\mathbf{n}} dS \tag{22}$$

For electromagnetic pulse sources, the surface integrals are zeros since we can always put the observation surface outside the region containing the fields. For harmonic waves, it can be proved that the surface integral at $S_\infty$ in (21) approaches zero, while that in (22) may be a nonzero but finite value[14].

For pulse radiators, the surface integral in (13) could be eliminated by choosing the observation surface outside the region $V_{rad}$. Therefore, the radiative power of the source can be evaluated in the source region,

$$P_{rad}(t) = \frac{\partial W_{rad}}{\partial t} = -\int_{V_s} \left[ \mathbf{E} \cdot \mathbf{J} + \frac{\partial}{\partial t} \left( \frac{1}{2} \rho \phi + \frac{1}{2} \mathbf{A} \cdot \mathbf{J} \right) \right] d\mathbf{r}' \tag{23}$$

$P_{rad}(t)$ describes the radiative power emitted by the source at the real time t. For pulse sources, it has nonzero values only in the time period $(0 \leq t \leq T)$. The variable $P_{Srad}(t)$ defined in (14) is the power passing the observation surface $S_a$.

It has nonzero values over period $(t_{\min} < t < t_{\max})$, which varies with the choice of the observation surface. Apparently, they are not expected to be equal. Only their integrals over the corresponding time duration are equal.

## IV. Radiation of Harmonic Sources

For harmonic fields with time convention of $e^{j\omega t}$, the radiation is assumed to last from $-\infty$ to $+\infty$, so the radiative energy is infinitely large. The Poynting theorem can be applied to describe the balance between the averaged powers and the varying rate of energies,

$$-\frac{1}{2}\int_{V_s} \mathbf{E}\cdot \mathbf{J}^* d\mathbf{r}' = 2j\omega \int_{V_a}\left[\frac{1}{4}\mathbf{B}^*\cdot \mathbf{H} - \frac{1}{4}\mathbf{E}\cdot \mathbf{D}^*\right]d\mathbf{r}' + \frac{1}{2}\oint_{S_a} \mathbf{E}\times \mathbf{H}^* \cdot \hat{\mathbf{n}} dS \quad (24)$$

From which the average radiative power at infinity can be evaluated with source distributions,

$$(P_{rad})_{av} = \text{Re}\left\{\frac{1}{2}\oint_{S_\infty} \mathbf{E}\times \mathbf{H}^* \cdot \hat{\mathbf{n}} dS\right\} = -\text{Re}\left\{\frac{1}{2}\int_{V_s} \mathbf{E}\cdot \mathbf{J}^* d\mathbf{r}'\right\} \quad (25)$$

However, the evaluation of the reactive energies in conventional formulations requires to subtract the radiative energy from the total energy. Since both the energies are unbounded values, all those formulations based on energy subtraction are not quite satisfactory so far.

With the theory proposed here, the power balance can be evaluated within any domain enclosed by an observation surface $S_a$ enclosing the source region $V_s$,

$$-\int_{V_s} \mathbf{E}\cdot \mathbf{J}^* d\mathbf{r}' = 2j\omega \int_{V_a}\left(\frac{1}{4}\phi\rho^* + \frac{1}{4}\mathbf{A}\cdot \mathbf{J}^*\right)d\mathbf{r}' + \oint_{S_a}\left[\frac{1}{2}\mathbf{E}\times \mathbf{H}^* - j\omega\left(\frac{1}{4}\mathbf{H}^*\times \mathbf{A} + \frac{1}{4}\phi\mathbf{D}^*\right)\right]\cdot \hat{\mathbf{n}} dS \quad (26)$$

The averaged radiative power crossing the observation surface can be obtained using the radiative power flux vector $\mathbf{S}_{rad}$ or the source distributions,

$$(P_{rad})_{av} = \text{Re}\left\{\oint_{S_a}\left[\frac{1}{2}\mathbf{E}\times \mathbf{H}^* - \frac{1}{4}j\omega(\mathbf{H}^*\times \mathbf{A} + \phi\mathbf{D}^*)\right]\cdot \hat{\mathbf{n}} dS\right\} = -\text{Re}\left\{\int_{V_s}\frac{1}{2}\mathbf{E}(\mathbf{r}')\cdot \mathbf{J}^*(\mathbf{r}')d\mathbf{r}'\right\} \quad (27)$$

Note that the observation surface is not required to approach infinity for evaluating the radiative power with the radiative power flux density vector. It can be checked that the result is in consistent with that obtained using the Poynting vector, as has been shown in [14] that

$$\text{Re}\left\{\oint_{S_\infty} j\omega\left(\frac{1}{4}\mathbf{H}^*\times \mathbf{A} + \frac{1}{4}\phi\mathbf{D}^*\right)\cdot \hat{\mathbf{n}} dS\right\} = 0 \quad (28)$$

The average reactive energy can be calculated with the source-potential products,

$$(W_{react})_{av} = \text{Re}\left\{\int_{V_s}\left(\frac{1}{4}\rho\phi^* + \frac{1}{4}\mathbf{A}\cdot \mathbf{J}^*\right)d\mathbf{r}'\right\} \quad (29)$$

Alternatively, making use of (21) and (22), the averaged reactive energy can be calculated using the fields and the vector potentials,

$$(W_{react})_{av} = \text{Re}\left\{\int_{V_\infty}\left(\frac{1}{4}\mathbf{E}\cdot \mathbf{D}^* + \frac{1}{4}\mathbf{B}\cdot \mathbf{H}^* + \frac{1}{2}j\omega\mathbf{A}\cdot \mathbf{D}^*\right)\right\} \quad (30)$$

In this section, the same symbols are used for the corresponding phasors for the sake of convenience.

## V. Hertzian Dipole

A Hertzian dipole locating at the origin is analyzed to show the energy/power balance relationship. The moment of the dipole is assumed to be $ql\cos\omega t$, the scalar potential and the vector potential of which can be readily derived from a Hertzian potential $\Pi = (ql/4\pi r)\cos(\omega t - kr)$ [44][45],

$$\mathbf{A} = -\frac{\omega\mu_0 ql}{4\pi r}\sin(\omega t - kr)\left(\hat{\mathbf{r}}\cos\theta - \hat{\boldsymbol{\theta}}\sin\theta\right) \quad (31)$$

$$\varphi = \frac{ql}{4\pi\varepsilon_0}\cos\theta\left[\frac{1}{r^2}\cos(\omega t - kr) - \frac{k}{r}\sin(\omega t - kr)\right] \quad (32)$$

from which the fields are found to be

$$\mathbf{E} = \frac{k^2 ql}{4\pi\varepsilon_0 r} \left\{ \hat{\mathbf{r}} 2\cos\theta \left[ \frac{1}{k^2 r^2} \cos(\omega t - kr) - \frac{1}{kr}\sin(\omega t - kr) \right] + \hat{\boldsymbol{\theta}}\sin\theta \left[ \left(\frac{1}{k^2 r^2} - 1\right)\cos(\omega t - kr) - \frac{1}{kr}\sin(\omega t - kr) \right] \right\} \quad (33)$$

$$\mathbf{H} = -\frac{\omega k q l}{4\pi r}\sin\theta \left[ \frac{1}{kr}\sin(\omega t - kr) + \cos(\omega t - kr) \right] \hat{\boldsymbol{\phi}} \quad (34)$$

The total reactive energies at the instant of time t stored in the whole space outside a small sphere with radius a can be derived from the fields and potentials using (21) and (22). The integrations in (22) are performed to get

$$\int_{V_\infty - V_a} \left( \frac{1}{2}\mathbf{B}\cdot\mathbf{H} - \frac{1}{2}\mathbf{A}\cdot\frac{\partial \mathbf{D}}{\partial t} \right) d\mathbf{r}' = \alpha_0 \left[ \frac{1}{ka} - \frac{1}{ka}\cos 2(\omega t - ka) + \sin 2(\omega t - ka) \right] - \alpha_0 \lim_{r\to\infty}\sin 2(\omega t - kr) \quad (35)$$

$$\oint_{S_\infty} \left( \frac{1}{2}\mathbf{H}\times\mathbf{A} \right)\cdot\hat{\mathbf{n}} dS = \alpha_0 \lim_{r\to\infty}\sin 2(\omega t - kr) \quad (36)$$

where $\alpha_0 = (\omega q l)^2 \mu_0 k / (24\pi)$. Combing (35) and (36) gives the stored reactive magnetic energy

$$W_m(t) = \alpha_0 \left[ \frac{1}{ka} - \frac{1}{ka}\cos 2(\omega t - ka) + \sin 2(\omega t - ka) \right] \quad (37)$$

The reactive electric energy can be calculated in a similar way,

$$W_e(t) = \alpha_0 \left[ \frac{1}{k^3 a^3} + \frac{1}{ka} + \left(\frac{1}{k^3 a^3} - \frac{1}{ka}\right)\cos 2(\omega t - ka) - \frac{2}{k^2 a^2}\sin 2(\omega t - ka) \right] \quad (38)$$

Note that the surface integral in (21) is zero.

The radiative power evaluated at the surface of the small sphere is

$$P_{rad}(t) = \oint_{S_a} \mathbf{S}_{rad}\cdot\hat{\mathbf{n}} dS = 2\omega\alpha_0 \quad (39)$$

It is a constant value independent of the radius of the sphere. The total power crossing any concentric spherical surface is the same.

For comparison, the surface integral of the Poynting vector on the spherical surface $S_a$ is

$$P_{pv}(t) = \oint_{S_a} \mathbf{S}\cdot\hat{\mathbf{n}} dS = 2\omega\alpha_0 \left[ 1 + \left(\frac{2}{ka} - \frac{1}{k^3 a^3}\right)\sin 2(\omega t - ka) + \left(1 - \frac{2}{k^2 a^2}\right)\cos 2(\omega t - ka) \right] \quad (40)$$

which varies with the radius of the surface. As expected, the time average of $P_{pv}(t)$ equals that of $P_{rad}(t)$. The time averaged energies are listed below,

$$\begin{cases} (W_m)_{av} = \alpha_0 \left(\frac{1}{ka}\right) \\ (W_e)_{av} = \alpha_0 \left(\frac{1}{k^3 a^3} + \frac{1}{ka}\right) \end{cases} \quad (41)$$

The Q factor of the dipole is then calculated to be

$$Q = \frac{2\omega(W_e)_{av}}{(P_{rad})_{av}} = \frac{1}{k^3 a^3} + \frac{1}{ka} \quad (42)$$

which is exactly in agreement with the result shown in [46].

The well-established equivalent circuit model proposed by Chu [47] for Hertzian dipole is shown in Fig.2. Assuming that the current in the radiation resistor at the interface of $r = a$ is $i_R = I_0 \cos(\omega t - ka)$, the energies stored in the capacitor and the inductor can be derived to be

$$\begin{cases} W_C = \frac{I_0^2}{2\omega}\left[ \frac{1}{ka} + \frac{1}{(ka)^3} + \left(\frac{1}{(ka)^3} - \frac{1}{ka}\right)\cos 2(\omega t - ka) - \frac{1}{(ka)^2}\sin 2(\omega t - ka) \right] \\ W_L = \frac{I_0^2}{2\omega}\left(\frac{1}{ka}\right)\sin^2(\omega t - ka) \end{cases} \quad (43)$$

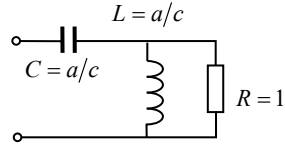

Fig.2 Equivalent circuit model for Hertzian dipole radiation.

If we choose $I_0^2 = 2\omega\alpha_0$, it can be checked that $W_C(t) = W_e(t)$, and $W_L(t) = W_m(t)$. This exact agreement gives a good support to the proposed theory.

## VI. Short Pulse Radiators

Two simple but typical radiators are analyzed to further support the theory. The first one is a short pulse source on a square patch. The second example is a solenoidal loop current. Their initial and final reactive energies are set to be zero.

In the examples, two spherical surfaces are chosen as observation surface, with their centers coinciding with that of the source. They are labeled as sphere-1 and sphere-2, with radius of 0.2m and 10m, respectively. The radiative energies $W_{rad}(t)$ passing through sphere-1, 2 are calculated with the second method. $W_{pv}(t)$ is the integration of the Poynting vector power passing through the observation surfaces,

$$W_{pv}(t) = \int_0^t P_{Spv}(\tau)d\tau = \int_0^t \oint_{sphere-1,2} \mathbf{S}(\mathbf{r}',\tau)\cdot\hat{\mathbf{n}}d\mathbf{r}'d\tau \qquad (44)$$

### A. Short pulse square patch radiator

The surface source resides on a square plate consisting of two connected triangles $Tr^\pm$ sharing a common edge with length of 0.2m, as shown in Fig 3(a).

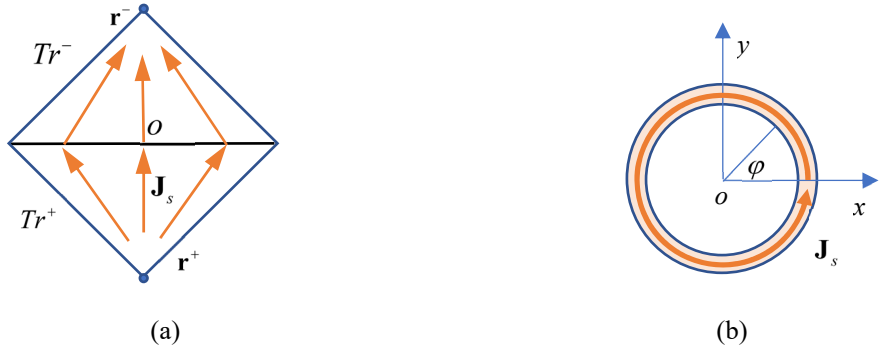

(a)          (b)

Fig. 3 Two sources. (a) Square patch source. (b) Solenoidal loop current.

Assume that the surface current density can be expressed with the product of a spatial function and a temporal function, $\mathbf{J}_s(\mathbf{r},t) = \mathbf{f}(\mathbf{r})I(t)$, with

$$\mathbf{f}(\mathbf{r}) = \begin{cases} \mathbf{r} - \mathbf{r}^+, \mathbf{r} \in Tr^+ \\ \mathbf{r}^- - \mathbf{r}, \mathbf{r} \in Tr^- \end{cases} \qquad (45)$$

The surface charge density has to observe the charge conservation law, namely, $\nabla_s \cdot \mathbf{J}_s + \partial\rho_s/\partial t = 0$, hence,

$$\rho_s(\mathbf{r},t) = -\int_{-\infty}^t \nabla_s \cdot \mathbf{J}_s(\mathbf{r},\tau)d\tau = \nabla_s \cdot \mathbf{f}(\mathbf{r})\int_{-\infty}^t I(\tau)d\tau \qquad (46)$$

Expressing the surface charge density as $\rho_s(\mathbf{r},t) = \Lambda(\mathbf{r})q(t)$,

$$\begin{cases} \Lambda(\mathbf{r}) = \nabla_s \cdot \mathbf{f}(\mathbf{r}) \\ q(t) = -\int_{-\infty}^t I(\tau)d\tau \end{cases} \qquad (47)$$

Consider a smooth pulse source that exists in the duration $0 \le t \le T$. To ensure a null initial and final state reactive energy, we choose,

$$q(t) = \begin{cases} (1-\cos\omega t)^2/\omega, 0 \le t \le T \\ 0, \text{else} \end{cases} \qquad (48)$$

and $\omega = 2\pi \times 10^9$, $T = 1\text{ns}$. The excitation energy $W_{exc}(t)$, the radiative energy $W_{rad}(t)$ and the reactive energy $W_{react}(t)$ associated with the source are shown in Fig. 4(a). The excitation energy is completely transformed to the radiative energy at $t = T$ since the initial and final reactive energy are zero. The energies passing through sphere-1 are shown in Fig. 4(b). The smallest and the largest distance between the source and sphere-1 are respectively 0.1m and 0.3m. Therefore, the radiative fields start to cross sphere-1 at about 0.33ns and end passing at about 2ns. The total radiative energy passed spherer-1 till t=2ns accurately equals the total excitation energy. Although in this case $W_{pv}(t)$ at t=2ns is also of the same level, it does not coincide with $W_{rad}(t)$ at other instant of time. The radiative fields pass sphere-2 in a similar manner as shown in Fig. 4(c). Due to the triangular mesh errors in the calculation, the time window for the radiative fields to pass sphere-2 is about 32.5ns<t<34.5ns, slightly different from that of an ideal spherical surface.

The excitation power, radiative power and the time varying rate of the reactive energy are shown in Fig. 5(a). It can be seen that in period $0 < t < t_1$, the excitation source contributes to the radiative power and the increment of the reactive energy; when $t_1 < t < t_2$, the reactive energy begins to decrease. The radiative power is caused by the excitation and the decreasing of the reactive energy; while in period $t_2 < t < T$, the reactive energy gradually decreases to zero. A part of the reactive energy is transformed to the radiative power, while the left part is absorbed by the sources, causing a negative excitation power.

As pointed out previously, the sources may couple with the surrounding fields and may have two consequences: change the reactive energy and/or excite additional radiative fields. In classical charged particle theory, it is known that the decreasing of the Schott energy, which is responsible for self energy, may cause radiation. An intuitive example is the system consisting of two massless particles with equal but opposite charges. In the first stage, the two particles are dragged away acceleratively from each other by an external force. The reactive energy of the system will increase and radiative fields will appear. In the second stage, remove the external force and let the two particles return to the initial state completely by the electromagnetic force between them. The reactive energy will decrease to the initial level, and in the meantime the system also will excite radiative fields due to the acceleration by their mutual force, accompanying with the decreasing of the self energy.

The Poynting vector consists of the radiative power flux density and the fluctuation power density. Their surface integrations of the three vectors on sphere-2 are shown in Fig. 5(b), where the fluctuation power is,

$$P_{Sreact}(t) = \oint_{S_a} \mathbf{S}_{react}(\mathbf{r}',t) \cdot \hat{\mathbf{n}} d\mathbf{r}' \tag{49}$$

The radiative power $P_{Srad}(t)$ varies smoothly and remains positive. The fluctuation power $P_{Sreact}(t)$ oscillates and can be negative, which clearly reflects the vibration of the reactive energy. Owing to its influence, the Poynting vector power $P_{Spv}(t)$ also shows ripples in its curve.

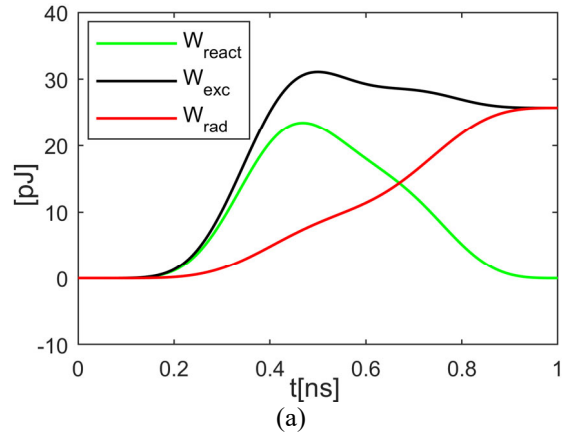

(a)

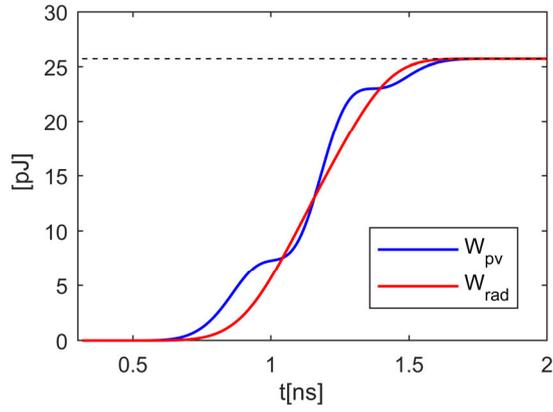

(b)

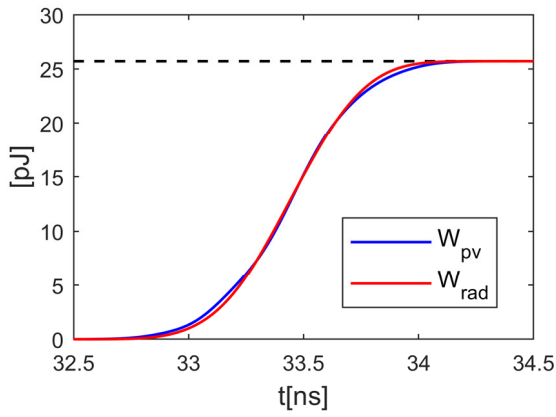

(c)

Fig.4 The energies of the square patch sources. (a) The total excitation energy, reactive energy and the radiative energy evaluated in the source region. (b) The radiative energy crossing sphere-1. (c) Radiative energy crossing sphere-2.

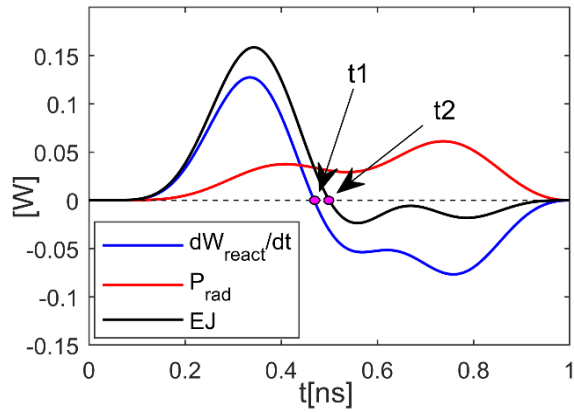

(a)

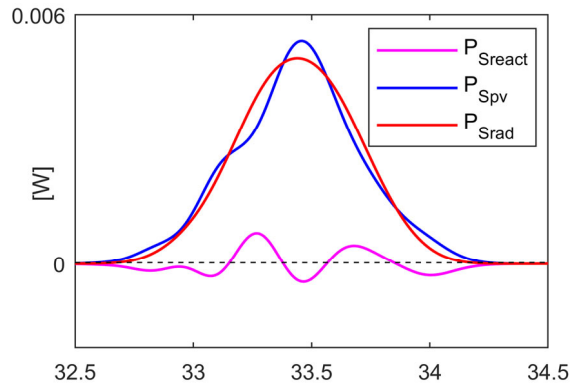

(b)

Fig.5 The powers of the square patch sources. (a) The excitation power, radiative power and the varying rate of the reactive energy evaluated in the source region.   (b) The radiative power, fluctuation power and the Poynting power crossing sphere-2.

### B. Loop radiator with solenoidal current

The solenoidal surface current on a ring is also described by $\mathbf{J}_s(\mathbf{r},t) = \mathbf{f}(\mathbf{r})I(t)$, as shown in Fig.3(b). Here,

$$\mathbf{f}(\mathbf{r}) = 1.0\hat{\varphi} \tag{50}$$

The inner and outer radius of the ring is 0.08m and 0.1m, respectively. The temporal function is a modulated Gaussian pulse,

$$I(t) = \begin{cases} e^{-\gamma^2}\sin\omega t, & 0 \leq t \leq T \\ 0, & \text{else} \end{cases} \tag{51}$$

with $\omega = 2\pi \times 10^{10}$, $\gamma = 2\sqrt{5}(t - 0.5T)/T$, $T = 1$ns.

The excitation energy, radiative energy and the energy evaluated with Poynting vector are shown in Fig. 6(a). The reactive energy includes the contribution from the current alone. It oscillates with the source and admits negative values periodically. In the proposed theory, it is acceptable because the reactive energy is dependent on the potentials, which are values relative to their reference zero points. Note that the Schott energy in the charged particle theory may also possibly be negative [19][48].

In this example, $\partial \rho_s / \partial t = -\nabla_s \cdot \mathbf{J}_s = 0$. The charge density corresponds to the loop current is a static one and has no influence on the radiation fields. On the other hand, we have $\mathbf{J}_s = \rho_s \mathbf{v}$. The charge density cannot be uniquely determined since the velocity is unknown. However, the static charge $\rho_s$ must satisfy $|\rho_s| \geq |\mathbf{J}_s|_{\max}/c = 1/c$ so as that the velocity of the charges is less than the light velocity at every instant of time. The static electric energy associated with the charge density is calculated to be $W_e \geq 1.12 \times 10^{-10}$ J, larger than the minimum reactive magnetic energy ($-3.1 \times 10^{-11}$ J).

The energies passing through sphere-1 are shown in Fig. 6(b).   The smallest and the largest distance between the source and sphere-1 are again respectively 0.1m and 0.3m. The total passed radiative energy at t=2ns is exactly of the same level as that evaluated at the source region.

The excitation power, radiative power and the time varying rate of the reactive energy are shown in Fig. 7(a). The reactive energy seems to perform like a bump, assisting the pulsed excitation to generate a smooth radiation.

The powers crossing sphere-2 are shown in Fig.7(b). The radiative power $P_{Srad}(t)$ varies smoothly and remains positive. The Poynting power contains ripples coming from the fluctuation power $P_{Sreact}(t)$.

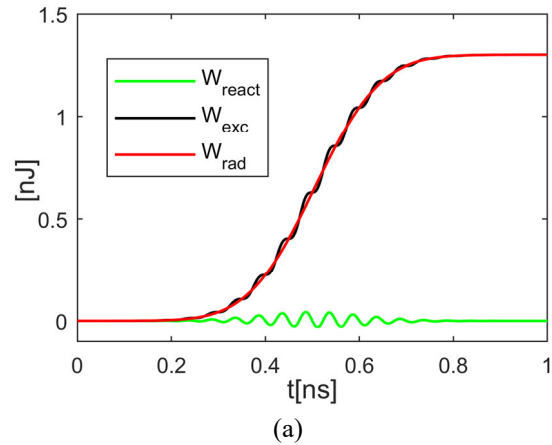

(a)

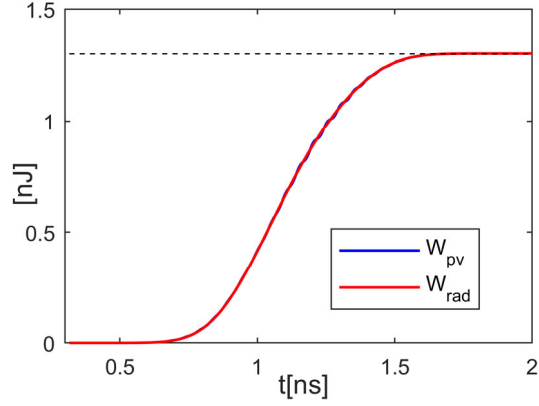

(b)

Fig.6 The energies of the loop current. (a) The total excitation energy, reactive energy and the radiative energy evaluated in the source region. (b) The radiative energy crossing sphere-1.

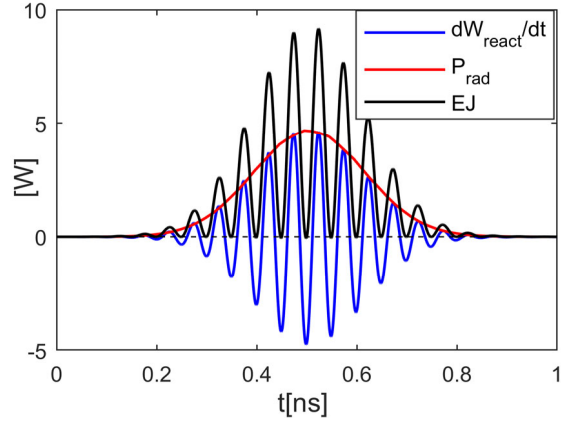

(a)

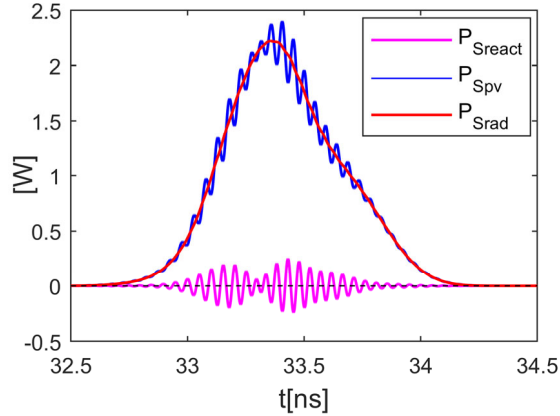

(b)

Fig.7 The powers of the loop current. (a) The excitation power, radiative power and the varying rate of the reactive energy evaluated in the source region. (b) The radiative power, fluctuation power and the Poynting power crossing sphere-2.

## VII. Conclusions

By reorganizing the power balance equation involved in electromagnetic radiation problems, a new vector for radiative power flux density is introduced. The proposed theory clearly reveals that the Poynting vector includes not only the radiative power flux density but also the influence of the fluctuation of the reactive energy. The theory is directly based on the macroscopic Maxwell equation and is non relativistic at its present formulation. Effort is making to get its correct relativistic counterpart.

It is true that the ideas discussed in this paper is slightly different from the traditional ones. The first is concerning with the Poynting vector. The proposed theory does not question the correctness of the Poynting Theorem, but argues that the interpretation of the Poynting vector and the way to apply the Poynting Theorem may be not satisfactory. The proposed theory can at least provide a better insight to the radiation problem. The second issue is about the reactive energy. The proposed definition can indeed provide a better interpretation to the radiation problem. At some instant of time, the excitation power evaluated in the source region may become negative and the radiator turns to an absorber. This is acceptable as we can address the radiator as a one-port device, and the reflected power may be larger than the input power at some times.

Although the issues discussed here is in free space, there seems to have no obvious barrier to prevent extending it to media environment in our future work.

# References

[1] R. E. Collin and S. Rothschild, "Evaluation of antenna Q," IEEE Trans. Antennas Propag., vol. AP-12, no. 1, pp. 23–27, Jan. 1964.
[2] A. Shlivinski and E. Heyman, "Time-domain near-field analysis of short pulse antennas—Part I: Spherical wave (multipole) expansion," IEEE Trans. Antennas Propag., vol. 47, no. 2, pp. 271–279, Feb. 1999.
[3] A. Shlivinski and E. Heyman, "Time-domain near-field analysis of short pulse antennas—Part II: Reactive energy and the antenna Q," IEEE Trans. Antennas Propag., vol. 47, no. 2, pp. 280–286, Feb. 1999.
[4] A. D. Yaghjian, "Internal energy, Q-energy, Poynting's theorem, and the stress dyadic in dispersive material," IEEE Trans. Antennas Propag., vol. 55, no. 6, pp. 1495–1505, Jun. 2007.
[5] G. A. E. Vandenbosch, "Reactive energies, impedance, and Q factor of radiating structures," IEEE Trans. Antennas Propag., vol. 58, no. 4, pp. 1112–1127, Apr. 2010.
[6] G. A. E. Vandenbosch, "Radiators in time domain—Part I: Electric, magnetic, and radiated energies," IEEE Trans. Antennas Propag., vol. 61, no. 8, pp. 3995–4003, Aug. 2013.
[7] G. A. E. Vandenbosch, "Radiators in time domain—Part II: Finite pulses, sinusoidal regime and Q factor," IEEE Trans. Antennas Propag., vol. 61, no. 8, pp. 4004–4012, Aug. 2013.
[8] M. Capek, L. Jelinek, P. Hazdra, and J. Eichler, "The measurable Q factor and observable energies of radiating structures," IEEE Trans. Antennas Propag., vol. 62, no. 1, pp. 311–318, Jan. 2014.
[9] M. Gustafsson and B. L. G. Jonsson, "Antenna Q and stored energy expressed in the fields, currents, and input impedance," IEEE Trans. Antennas Propag., vol. 63, no. 1, pp. 240–249, Jan. 2015.
[10] W. Geyi, "Stored energies and radiation Q," IEEE Trans. Antennas Propag., vol. 63, no. 2, pp. 636–645, Feb. 2015.
[11] M. Capek, L. Jelinek, and G. A. E. Vandenbosch, "Stored electromagnetic energy and quality factor of radiating structures," Proc. Roy. Soc. A, Math., Phys. Eng. Sci., vol. 472, no. 2188, pp. 20150870, 2016.
[12] G. A. E. Vandenbosch, "Recoverable energy of radiating structures," IEEE Trans. Antennas Propag., vol. 65, no. 7, pp. 3575–3588, Jul. 2017.
[13] K. Schab et al., "Energy stored by radiating systems," IEEE Access, vol. 6, pp. 10553–10568, 2018.
[14] G. B. Xiao, C. Xiong, S. Huang, R. Liu, Y. Hu, "A new perspective on the reactive electromagnetic energies and Q factors of antennas," IEEE Access, vol. 8, 8999565, pp. 173790-173803, Oct. 2020.
[15] J. D. Jackson, "Classical Electrodynamics;" 3rd ed., John Wiley & Sons: New York, NY, USA, 1998.
[16] F. Rohrlich, "Classical Charged Particles," 3rd ed., World Scientific Publishing: Singapore, 2007.
[17] T. Nakamura, "On the Schott term in the Lorentz-Abraham-Dirac equation," Quantum Beam Sci., vol. 4, pp. 34, 2020.
[18] G. A. Schott, "Electromagnetic radiation and the mechanical reactions arising from it," Cambridge University Press: Cambridge, UK, 1912.
[19] Ø. Grøn, "The significance of the Schott energy for energy-momentum conservation of a radiating charge obeying the Lorentz-Abraham-Dirac equation," Am. J. Phys., vol. 79, no. 1, pp. 115–122, 2011.
[20] J. H. Poynting, "On the connexion between electric current and the electric and magnetic inductions in the surrounding field," Proc. Royal Soc. London. vol. 38, pp. 168-172, 1884-1885.
[21] C. S. Lai, "Alternative choice for the energy flow vector of the electromagnetic field," Am. J. Phys., vol. 49, no.9, 841-843, Jan. 1981.
[22] R. H. Romer, "Alternatives to the Poynting vector for describing the flow of electromagnetic energy," Am. J. Phys., vol. 50, no.12, pp.1166-1168, Nov. 1982.
[23] F. Herrmann, and G. Bruno Schmid, "The Poynting vector field and the energy flow within a transformer," Am. J. Phys., vol. 54, no.6, 528-531, Jun. 1986.
[24] C. J. Carpenter, "Electromagnetic energy and power in terms of charges and potentials instead of fields." IEE Proc. A, vol. 136, no. 2, pp.55-65, Mar 1989.
[25] J. A. Ferreira, "Application of the Poynting vector for power conditioning and conversion," IEEE Trans. Education, vol. 31. no. 4, pp. 257-264, Nov. 1988.
[26] D. F. Nelson, "Generalizing the Poynting vector," Physical Rev. Lett., vol. 76, no. 25, 17, Jun. 1996.
[27] W. Gough, "Poynting in the wrong direction?" Eur. J. Phys., vol.3, no.2, pp.83-87, Dec. 2000.

[28] L. S. Czarnecki, "Energy flow and power phenomena in electrical circuits: illusions and reality," Electrical Engineering, vol. 82, no.3, pp.119-126, Mar. 2000.
[29] Z. Cakareski, A. E. Emanuel, "Poynting vector and the quality of electric energy," European Trans. Electrical Power, vol. 11, no. 6, 375-381, Nov. 2001.
[30] A. Chubykaloa, A. Espinozab, and R. Tzonchevc, "Experimental test of the compatibility of the definitions of the electromagnetic energy density and the Poynting vector," European Physical J. D, vol. 31, no.1, pp.113–120, Oct. 2004.
[31] L. S. Czarnecki, "Could power properties of three-phase systems be described in terms of the Poynting vector?" IEEE Trans. Power Delivery, vol. 21, no. 1, pp. 339-344, Jan. 2006.
[32] A. E. Emanuel, "Poynting vector and the physical meaning of nonactive powers," IEEE Trans. Instrument. Measure., vol. 54, no. 4, pp.1457-1462, Aug. 2005.
[33] A.E. Emanuel, "About the rejection of Poynting vector in power systems analysis," J. Electrical Power quality Utilization, vol.8, no.1, pp.43-48, 2007.
[34] P. Kinsler, A. Favaro and M. W. McCall, "Four Poynting theorems," Eur. J. Phys., vol. 30, no. 5, pp. 983–993, Aug. 2009.
[35] J. D. Jackson, "How an antenna launches its input power into radiation: the pattern of the Poynting vector at and near an antenna," Am. J. Phys., vol. 74, no.4, 280-288, Jul. 2005.
[36] E. K. Miller, "The differentiated on-surface pointing vector as a measure of radiation loss from wires," IEEE Antennas Propagat. Magazine, vol. 48, no. 6, pp.21-32, Dec. 2006.
[37] U. Lundin, B. Bolund, and M. Leijon, "Poynting vector analysis of synchronous generators using field simulations," IEEE Trans. Magn., vol. 43, no. 9, pp. 3601-3607, Sep. 2007.
[38] I. Mokhun, A. Arkhelyuk, Y. Galushko, Y. Kharitonovtta, and J. Viktorovskaya, "Experimental analysis of the Poynting vector characteristics," Applied Optics, vol. 51, no. 10, pp.158-161, Apr. 2012.
[39] K. Cheng, X. Zhong, A. Xiang. "Propagation dynamics, Poynting vector and accelerating vortices of a focused Airy vortex beam," Optics Laser Techn., vol. 57, pp. 77–83, Apr. 2014.
[40] M. I. Marqués, "Beam configuration proposal to verify that scattering forces come from the orbital part of the Poynting vector," Optics Lett., vol. 39, no. 17, pp. 5122-5125, Sep. 2014.
[41] A. Kholmetskii, O. Missevitch, T. Yarman, "Poynting Theorem, relativistic transformation of total energy–momentum and electro- magnetic energy–momentum tensor," Found Phys., vol. 46, pp. 236-261, 2016.
[42] A. K. Singal, "Poynting flux in the neighborhood of a point charge in arbitrary motion and radiative power losses," Eur. J. Phys., vol. 37 045210, May 2016.
[43] Xuezhe Tian, Gaobiao Xiao, and Shang Xiang, "Application of analytical expressions for retarded-time potentials in analyzing the transient scattering by dielectric objects," IEEE Antennas Wireless Propag. Lett., vol.13, pp.1313-1316, 2014.
[44] G. B. Xiao, "Electromagnetic energy balance equations and Poynting Theorem," arXiv: 1910.02468v3 [physics.class-ph].
[45] J. A. Kong, "Fundamentals in electromagnetic wave theory," 3rd ed., Cambridge, MA, USA: EMW Publishing, 2008, pp.65-67.
[46] J. S. McLean, "A re-examination of the fundamental limits on the radiation Q of electrically small antennas," IEEE Trans. Antennas Propag., vol. 44, no. 5, pp. 672–676, May 1996.
[47] L. J. Chu, "Physical limitations on omni-directional antennas," J. Appl. Phys., vol. 19, no. 12, pp. 1163–1175, 1948.
[48] D. R. Rowland, "Physical interpretation of the Schott energy of an accelerating point charge and the question of whether a uniformly accelerating charge radiates," Eur. J. Phys., vol. pp. 31, 1037–1051, Jul. 2010.